\documentclass[aps,superscriptaddress,prl,twocolumn]{revtex4-1}
\usepackage{graphicx}
\usepackage{epstopdf}
\usepackage{array}
\usepackage{color}
\usepackage{amsmath}
\usepackage{amsxtra}
\usepackage{amstext}
\usepackage{amssymb}
\usepackage{latexsym}
\usepackage{float}
\usepackage{color}
\usepackage[colorlinks=true, linkcolor=blue,urlcolor=blue,citecolor=blue]{hyperref}

\begin{document}

\title{Competing Fractional Quantum Hall and Electron Solid Phases in Graphene}
\author{Shaowen Chen}
\thanks{S.Ch. and R. R.-P. contributed equally to this work}
\affiliation{Department of Physics, Columbia University, New York, NY, USA}
\affiliation{Department of Applied Physics and Applied Mathematics, Columbia University, New York, NY, USA}

\author{Rebeca Ribeiro-Palau}

\altaffiliation[Present address: ]{Centre de Nanosciences et de Nanotechnologies (C2N), CNRS, Univ Paris Sud,
Universit\'e Paris-Saclay, 91120 Palaiseau, France}
\email{rebeca.ribeiro@c2n.upsaclay.fr}
\affiliation{Department of Physics, Columbia University, New York, NY, USA}
\affiliation{Department of Mechanical Engineering, Columbia University, New York, NY, USA}

\author{Kang Yang}
\affiliation{Laboratoire de Physique des Solides, CNRS UMR 8502, Universit\'e Paris-Sud, Universit\'e Paris Saclay, 91405 Orsay cedex 91405, France}
\affiliation{LPTHE, CNRS-Universit\'e Pierre et Marie Curie, Sorbonne Universit\'es, 4 Place Jussieu, 75252 Paris Cedex 05, France}
\author{Kenji Watanabe}
\affiliation{National Institute for Materials Science, 1-1 Namiki, Tsukuba, Japan}
\author{Takashi Taniguchi}
\affiliation{National Institute for Materials Science, 1-1 Namiki, Tsukuba, Japan}
\author{James Hone}
\affiliation{Department of Mechanical Engineering, Columbia University, New York, NY, USA}
\author{Mark O. Goerbig}
\affiliation{Laboratoire de Physique des Solides, CNRS UMR 8502, Universit\'e Paris-Sud, Universit\'e Paris Saclay, 91405 Orsay cedex 91405, France}
\author{Cory R. Dean}
\email{cd2478@columbia.edu}
\affiliation{Department of Physics, Columbia University, New York, NY, USA}

\begin{abstract}
We report experimental observation of the reentrant integer quantum Hall effect in graphene, appearing in  the N$=$2 Landau level.  Similar to high-mobility GaAs/AlGaAs heterostructures, the effect is due to a competition between incompressible fractional quantum Hall states, and electron solid phases. 
The tunability of graphene allows us to measure the $B$-$T$ phase diagram of the electron-solid phase. The hierarchy of reentrant states suggests spin and valley degrees of freedom play a role in determining the ground state energy. We find that the melting temperature scales with magnetic field, and construct a phase diagram of the electron liquid-solid transition.

\end{abstract}

\maketitle

Electrons confined to two dimensions and subjected to strong magnetic fields can host a variety of fascinating correlated electron phases. One of the most widely studied examples is the fractional quantum Hall effect (FQHE)\cite{tsui1982two, laughlin1983anomalous,Haldane1983, jain1989composite}, an incompressible liquid that emerges when the lowest energy Landau levels (LLs) are partially filled. However, the incompressible FQHE liquids are not the only correlated phases that can emerge within partially filled LLs and generically compete with the formation of interaction-driven electron solids, such as the Wigner crystal \cite{Wigner1934,Monarkha2012,Drichko2016,Drichko2015}, and the bubble \cite{fogler1996ground,lewis2002microwave,goerbig2003microscopic,goerbig2004competition,gervais2004competition,deng2012contrasting,pan2014competing,Du1999,lilly1999evidence,Shingla2018} and stripe charge density wave states\cite{fogler1996ground,MC96,lilly1999evidence,goerbig2003microscopic,goerbig2004competition,Du1999}. 

In GaAs/AlGaAs heterostructures, the competition between these different phases,  particularly developed in the N = 1 and 2  LL, gives rise to  a reentrant integer quantum Hall effect (RIQHE) \cite{eisenstein2002insulating, xia2004electron,kumar2010nonconventional,li2010observation,deng2012contrasting}. This is characterized by the emergence of vanishing longitudinal resistance at fractional filling between the usual sequence of FQHE states, but with Hall conductivity restored to the closest integer value.
Numerous experimental\cite{lewis2002microwave,lewis2005microwave,deng2012contrasting,deng2012collective,wang2015depinning} and theoretical studies\cite{goerbig2003microscopic,goerbigRFQHE,goerbig2004competition} favor a collective origin for the RIQHE where the emergent electron solid is pinned by the underlying impurity potential. However, many of the experimentally reported details, such as the relative energy scales between different RIQHE states and apparent particle-hole asymmetry within a LL\cite{deng2012collective,deng2012contrasting} remain poorly understood.

The universality of the integer and fractional QHE found in a wide variety of high mobility 2D electron systems suggests that the formation of the electron solid should be equally ubiquitous.  However,  observation of the RIQHE has so far remained conspicuously limited to GaAs heterostructures. In this Letter, we report experimental observation of a RIQHE in monolayer graphene, appearing near 0.33 partial filling of the  N = 2  LL, together with weakly formed FQHE states at $1/5$ in this same LL. Our results are in excellent agreement with recent theoretical calculations that suggest that the solid phase is stabilized and dominates over the FQHE liquid in graphene at these filling fractions\cite{papic2011tunable,knoester2016electron,zhang2007wigner}. The wide tunability of the carrier density in graphene allows us to map the evolution in both magnetic field and temperature of four distinct RIQHE states appearing within the lower spin branch of the N = 2 LL. Comparing their melting temperatures reveals an unexpected hierarchy that is consistent with a residual spin and/or valley symmetry, indicating that the expanded degrees of freedom in graphene play a role.  For a single RIQHE we use the melting transition to construct the first  $B$-$T$ phase diagram of the bubble phase at fixed filling fraction.   

\begin{figure}[t!]
\includegraphics[width=1\linewidth]{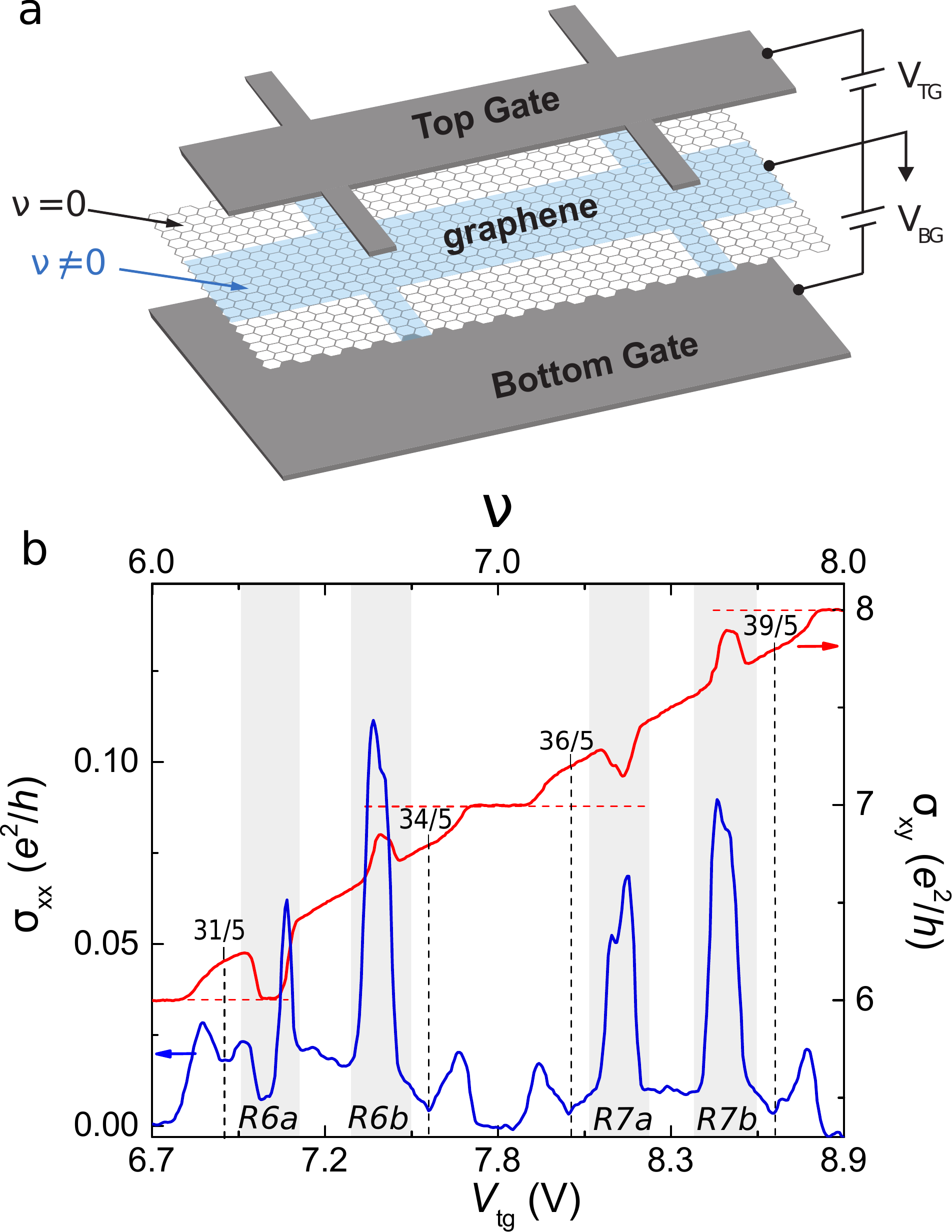} 
\caption{{\bf RIQHE in the N$=$2 LL of graphene.} {\bf a}, Schematics of the gate defined Hall bar device used in this experiment. {\bf b}, Longitudinal (left) and Hall (rigth) conductivities as a function of the filling factor measured at $B=23$~T and $T=0.3$~K. Colored regions highlight the four RIQHE and dashed vertical lines mark the four FQH states. Additionally, dashed horizontal lines mark the nearest integer value where the RIQHE is expected to quantized.}
\label{fig:fig1}
\end{figure}

Magnetoresistance measurements were performed in  electrostatically defined Hall bars of monolayer graphene \cite{Ribeiro2018}. The heterostructures were  prepared by a dry layer assembly technique with edge-contacts \cite{Wang_2013}, using graphite flakes as both top and bottom gates (Fig. \ref{fig:fig1}a). In brief (details of the device structure are given in \cite{Ribeiro2018}) the bottom gate spans the full width of the graphene layer whereas the top gate is etched into the shape of a Hall bar. The  bottom gate is biased such that the outer boundary of the device is maintained at the zero-density charge neutrality point: because the $\nu=0$ state in graphene state is gapped at moderate fields both in the bulk and edge\cite{young2012spin,young2014tunable} this acts as a depletion region. The top graphite gate then acts as an accumulation gate, and is used to define the carrier density away from $\nu=0$ in the Hall-bar shaped interior region of the device. Together this enables realization of a an electrostatically defined device (blue coloured region in Fig. \ref{fig:fig1}a). 

\begin{figure}[t!]
\includegraphics[width=1\linewidth]{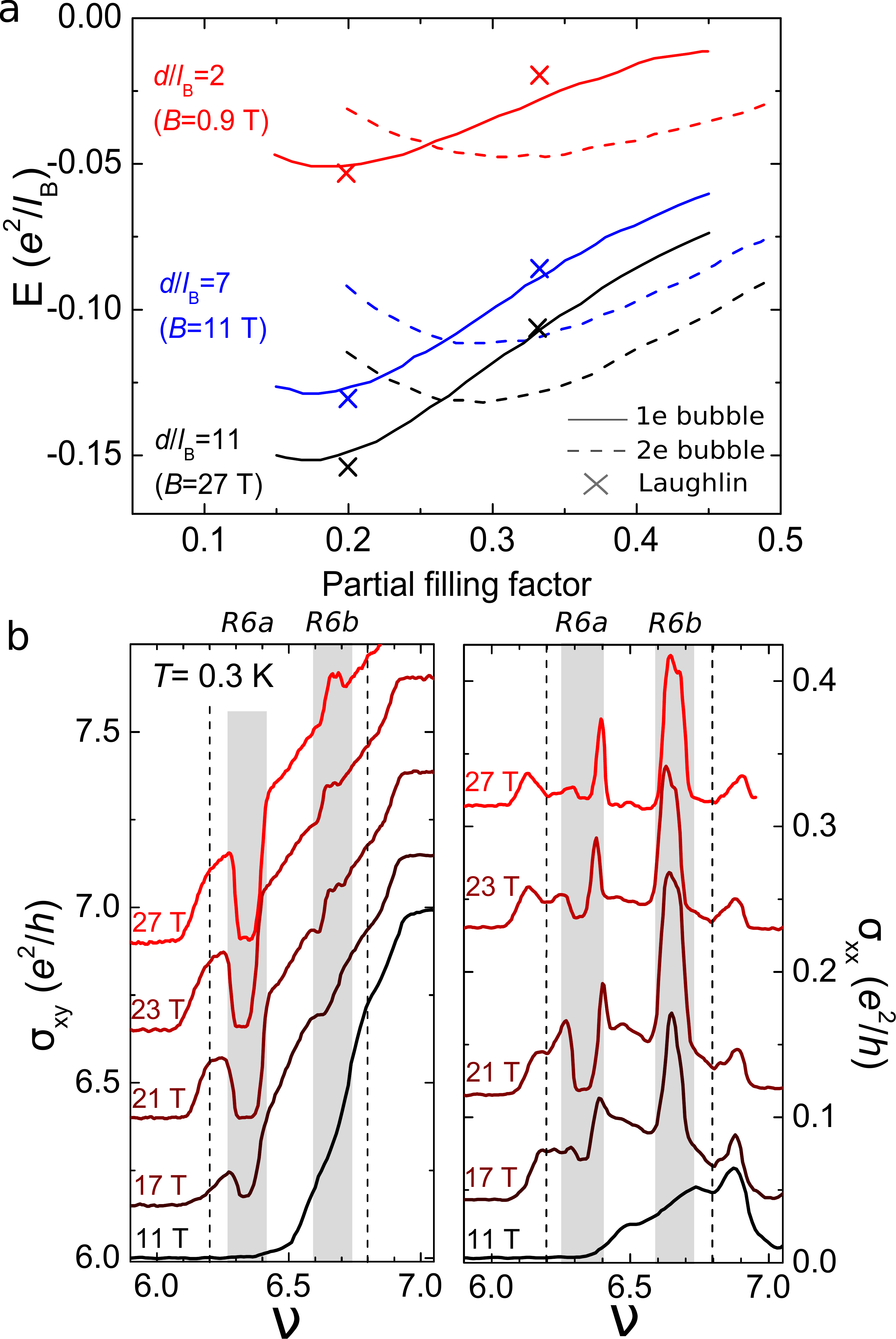}
\caption{ {\bf Magnetic field dependence of the electron solid phase.} {\bf a}  Numerical calculation of the energy as a function of filling for Laughlin and electron solid states in the N = 2 LL for different $d/l_{\rm B}$ ratios, where $d$ is the distance to the metallic gates.  The Laughlin states are represented by crosses while the one electron and two-electron bubble phases are represented by solid and dashed lines, respectively. {\bf b} Longitudinal (right) and Hall (left) conductivities as a function of filling factor for selected magnetic fields measured at 0.3 K. Dashed lines  mark the presence  of FQH states and colored regions highlight the two RIQHE. All curves have been shifted for clarity. } 
\label{fig:fig2}
\end{figure}

Figure \ref{fig:fig1}b shows the longitudinal ($\sigma_{\rm xx}$) and Hall ($\sigma_{\rm xy}$) conductivity for the lower spin branch of the N = 2 LL ($6<\nu< 8$), measured  at $B=23$~T and $T=0.3$~K. Four examples of the reentrant behaviour can be identified, which we label $R6a$ and $R6b$ for the first valley branch and $R7a$ and $R7b$ for the other valley branch. We note that only the $R6a$ state is fully developed, with $\sigma_{\rm xy}$  showing a quantized plateau at $6e^2/h$,  simultaneous with a well developed minimum in $\sigma_{\rm xx}$, where $e$ is the electron charge and $h$ is Planck's constant. For the remaining RIQHE states, where the Hall resistance is not fully quantized, the longitudinal conductivity shows a large local maximum, consistent with previous observations in GaAs\cite{eisenstein2002insulating,xia2004electron,deng2012contrasting,lewis2005microwave}.  In addition to the RIQHE states, we observe signatures of weakly developed FQHE states at $\nu=6+1/5$,  $6+4/5$, $7+1/5$, and  $7+4/5$ in the form of $\sigma_{\rm xx}$ minima simultaneous with kinks in the Hall conductivity (though not showing clear plateaus). Similar FQHE states have been previously reported in the third LL of ultra-high mobility GaAs/AlGaAs samples \cite{gervais2004competition}. Finally, we note that there is a clear absence of the $1/3$ FQHE states, which are the dominant FQHE states appearing in both N=0 and N=1 orbital branches of monolayer graphene\cite{deanFQHE10,amet2015composite}(also see supplementary information). Taken together these observations are in agreement with theoretical calculations indicating that charge density order is favoured over a Laughlin FQHE state at 1/3 filling in the N=2 LL, but the FQHE is favoured at 1/5 filling (Fig \ref{fig:fig2}a as well as Ref. \onlinecite{knoester2016electron}).

Fig. \ref{fig:fig2}a shows the theoretically calculated energy of the electron solid and FQHE states, for magnetic fields up to $B=27$~T. The energies of the crystalline phases have been calculated within the Hartree-Fock approximation, while those of the liquid FQH states have been obtained with the help of the plasma sum rules \cite{goerbig2004competition,knoester2016electron}.  In addition, we take into account explicitly the present experimental setup with metallic gates at a distance of 27 nm below and above  the graphene sheet. They screen the effective Coulomb interaction as a function of $d/l_{\rm B}$  (see supplementary information for further details), where $d$ is the distance between the gates, $l_{B}=\sqrt{\hbar/ eB}$ is the magnetic length. 

At all magnetic fields considered we find that the  electron solid is theoretically favourable over a FQHE at 1/3 filling, but the situation remains reversed at 1/5 filling with the FQHE state expected to be the ground state.  In Fig. 2b, we plot the evolution of the measured longitudinal and Hall conductivity in the filling factor range $\nu=$6 to 7, for magnetic field ranging from 11 T to 27 T. At $B=27$~T we observe a well developed $R6a$ but only weakly developed $R6b$ state, in addition to weakly formed FQHE states at 1/5 and 4/5 fillings, consistent with expectation. As we decrease the magnetic field the $R6b$ state quickly disappears. By contrast the more robust $R6a$ varies in width but remains well quantized at least down to 17~T.  The observed electron-hole asymmetry between the $R6a$ and $R6b$ is not anticipated by theoretical calculations, instead we expect these to be simple copies of each other \cite{poplavskyy,Goerbig2011grapheneRIQHE,knoester2016electron,zhang2007wigner}.

\begin{figure}[t]
\includegraphics[width=0.9\linewidth]{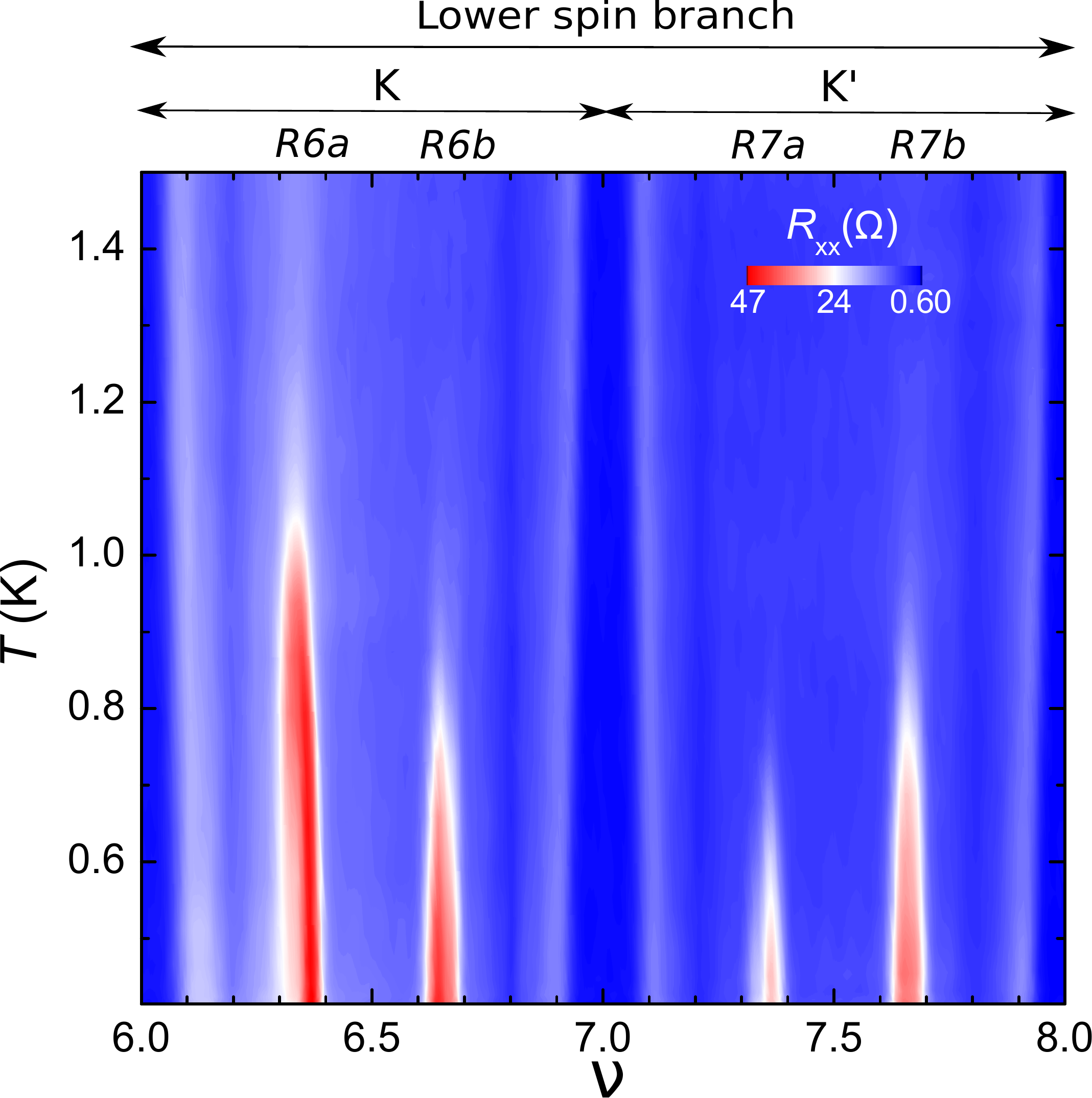}
\caption{{\bf Broken valley symmetry for the RIQHE of graphene}. {\bf a} Longitudinal resistance as a function of filling factor and  temperature, measured at 23 T, for the lower spin branch of the N = 2 LL.}
\label{fig:fig5}
\end{figure}

At lower magnetic fields, $B=11$ T, an unexpected behavior is observed near $R6a$, where the $\nu=6$ IQHE features, i.e. the Hall plateau and zero longitudinal conductivity, have become extended and merge with $R6a$ features. This same behaviour is not observed near the $\nu=7$ plateau, where instead, at this same field signatures of the 6+4/5 FQHE state remain and the $R6b$ features have simply disappeared giving way to an electron liquid. We interpret the extended $\nu=6$ plateau to indicate that in this field range, an electron solid state also exists near 1/5 filling. This is not in agreement with the calculations shown in Fig. \ref{fig:fig2}a and so the origin of this behavior is uncertain. We note that a similar transition of the electron solid as a function of magnetic field has been observed in the GaAs/AlGaAs quantum well system for the lowest LL when measured at different carrier densities \cite{liu2012observation}. 
In that study it was argued to be a quantum well width effect.  However this is unlikely in our case since electrons are confined to a single atomic layer.
We conjecture that in the low field limit an additional impurity potential may favor the electron solid phase over the FQHE near 1/5 filling. Indeed, allowing for local deformations of the lattice, an electron solid can profit more efficiently from the impurity potential than the incompressible FQH states \cite{goerbig2004competition}, theoretically inverting the relative ground state energies. The onset of this behavior in the low B limit could reflect a competition between the Coulomb and impurity energy scales.  We note also that we generally observe the reentrant state to become better developed with successive cool-downs (see Supplementary Information). Assuming that disorder increases after a thermal cycle, this observation would be consistent with disorder playing a role in stabilizing the RIQHE state.

\begin{figure*}[t!]
\includegraphics[width=1\linewidth]{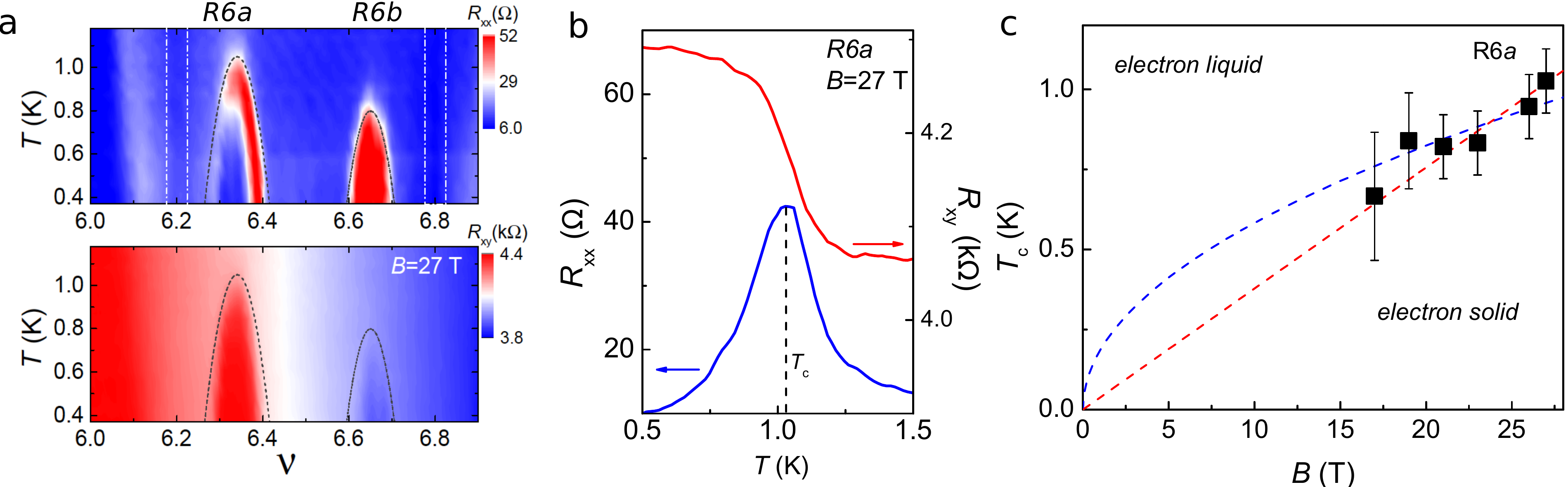}
\caption{{\bf Temperature dependence of the electron solid phase.} {\bf a} Hall and longitudinal resistance as a function of filling factor and 
temperature for the $R6a$ and $R6b$ reentrant states, measured at 27 T. White dashed boxes mark the FQH state $\nu= 6+ 1/5$ and $\nu=6 +4/5$ and gray dashed are fits using the parabolic function $T_c(\bar{\nu}) = T_c(\bar{\nu_c}) - \beta (\bar{\nu} - \bar{\nu_c})^2$ for each RIQHE, consistent with \cite{deng2012collective}. {\bf b}, line cut at $\nu=6.35$ for $R_{\rm xx}$ (left) and $R_{\rm xy}$ (right axis) at 27 T. {\bf c}, $B$-$T$ phase diagram of the $R6a$ state, dashed red line is the linear fit, dashed blue line is the square root fit, grey area indicates the electron solid regime.}
\label{fig:fig3}
\end{figure*}

Finally, we consider the temperature dependence of the RIQHE states. The critical temperature ($T_{\rm c}$),  where the electron solid undergoes a phase transition and melts into an electron liquid \cite{deng2012collective,deng2012contrasting}, provides a convenient estimate of the energy scale associated with the solid phase.
Fig. \ref{fig:fig5}a shows the longitudinal resistance versus temperature for filling fractions $\nu=$6 to 8, measured at $B=23$~T.  The four RIQHE states,  $R6a$, $R6b$, $R7a$ and $R7b$, identified in this plot by a resistive peak (red colour) appear to melt at different critical temperatures.  The qualitative trend in the apparent melting temperatures shows a relative hierarchy with $T_{c}^{R6a} > T_{c}^{R6b}$, while $T_{c}^{R7a} < T_{c}^{R7b}$.  This difference in the $a$ and $b$ instances of the RIQHE suggest that the two states are not related by electron-hole symmetry within a single spin-valley branch.  This result is unexpected\cite{deng2012collective}, 	since spin and valley degrees of freedom are not anticipated to play a role and the two RIQHE states are instead expected to be merely two spin copies of the same state, with identical melting temperatures.  We note that due to the elevated magnetic field (23 T) it is not possible for us to access the upper spin branch using the top gate. However, the hierarchy is suggestive, appearing symmetric at least across the entire spin branch.  This symmetry reflects a similar hierarchy identified in the FQHE states of the N = 1 LL of monolayer graphene\cite{feldman2012unconventional,zeng2018ultra,polshyn2018quantitative}, where it has been suggested that spin and valley degeneracies are only partially lifted, and an approximate SU(2) or SU(4) symmetry is preserved for the composite fermion ground states.  Our observation of a similar hierarchy in the RIQHE of the N = 2 LL suggests that similarly these degeneracies may be only partially lifted, and moreover, this order plays a role in the ground state energy of the solid phase as well.

Fig. \ref{fig:fig3}a,b shows high resolution maps of the temperature evolution of longitudinal and Hall resistance for the $R6a$ and $R6b$, acquired at $B=27$ T. The $R6a$ state is sufficiently well developed that we can quantitatively study its phase boundary. As the temperature is reduced, the resistance peak associated with $R6a$ state splits, while the reentrant Hall plateau grows wider in $\nu$.  The filling fraction boundary of both features follows an approximate parabolic shape, similar to what has been reported in GaAs\cite{deng2012collective}.

Fig. \ref{fig:fig3}b shows the temperature dependence of $R_{xx}$ and $R_{xy}$ acquired at $\nu=6.35$, where the melting temperature is a maximum.  As the temperature is lowered, $R_{\rm xx}$  increases to a maximum and then decreases, while $R_{\rm xy}$  increases from its classical Hall effect value to quantized 
integer quantum Hall level. Following Ref. \onlinecite{deng2012collective}, we take the temperature where $R_{\rm xx}$  is maximal to define the melting temperature, $T_{\rm c}$, of the solid phase.  At 27 T, the $R6a$ state has a $T_{\rm c}$  of about 1 K.

Extracting the melting temperature of $R6a$ at different magnetic fields, we have constructed a $B$-$T$ phase diagram, shown in Fig. \ref{fig:fig3}d. The melting temperature decreases as the magnetic field decreases. The electron solid state is driven by Coulomb interaction so a $\sqrt{B}$ dependence is expected \cite{deng2012collective}. On the other hand, a linear trend can also be expected due to screening from the top and back gates. This effect can be illustrated in the case of a single gate at a distance $d/2$ from the graphene layer, where the mirror charge creates a dipole and the interaction becomes dipolar at long distances (see supplementary information). The energy scale $E_c = e^2/4 \pi \epsilon l_{\rm B}\propto \sqrt{B}$ (Coulomb) thus needs to be roughly replaced by $ E_d = (e^2/4\pi\epsilon l_{\rm B})\times (d/l_{\rm B})\propto d \times B$ (dipolar). Naturally, one expects the effect to saturate in the large $d/l_{\rm B}$ limit, where the dipolar interaction is no longer justified. However, our measurement is made in an intermediate regime, where this expansion remains well justified. Over the field range where we can resolve the RIQHE the linear and square root fit equally well and we are unable to discriminate these dependences. Further study is needed to resolve these differences.

In conclusion, we have observed RIQHE in the N = 2 LL in graphene. The magnetic field evolution of 
states suggests a crossover of the energy competition between electron liquid and solid states. The temperature dependence of the states indicates a surprising 
hierarchy between the RIQHE states consistent with an approximate SU(2) or SU(4) symmetry being preserved.
We have extracted the onset temperature and constructed the 
$B$-$T$ phase diagram of the electron solid state. Our work opens the door of RIQHE study in a new, tunable material system, which contributes to the understanding of 
electron solid state in quantum Hall systems.

\section{\label{sec:level1}ACKNOWLEDGEMENT}
We thank G\'abor Cs\'athy, J.I.A. Li, M. Yankowitz and S. Dietrich for helpful discussions. S. Chen is supported by the ARO under MURI W911NF-17-1-0323.  This research was supported by the NSF MRSEC programme through Columbia in the Center 
for Precision Assembly of Superstratic and Superatomic Solids (DMR-1420634). A portion of this work was performed at the National High Magnetic Field Laboratory, which is supported by National Science Foundation Cooperative Agreement No. DMR-1157490 and the State of Florida.

\bibliography{Ref_v4}


\clearpage




\newcommand{\bq}{{\bf q}}
\newcommand{\bp}{{\bf p}}
\newcommand{\br}{{\bf r}}
\newcommand{\bR}{{\bf R}}
\newcommand{\rhobar}{\bar{\rho}}
\newcommand{\nubar}{\bar{\nu}}

\renewcommand{\thefigure}{S\arabic{figure}}
\renewcommand{\thesubsection}{S\arabic{subsection}}
\renewcommand{\theequation}{S\arabic{equation}}
\setcounter{figure}{0} 
\setcounter{equation}{0}

\section*{Supplementary Information}

\section*{Detail of the simulations}

As in the references \cite{knoester2016electron} and \cite{goerbig2004competition}, we consider a competition between electron-solid phases (Wigner and bubble crystals) described conveniently within the  Hartree-Fock approximation and Laughlin liquids the energy of which can be obtained to great accuracy by the so-called plasma sum rules \cite{Levesque1984,Girvin1986}. While the electron solids, formed by electrons in the last partially occupied Landau level (LL), are pinned by the underlying disorder and therefore provide an insulating response  (apart from the quantized Hall conductance, which stems from the completely filled lower LLs or spin-valley subbranches of Landau levels), the Laughlin liquids display the fractional quantum Hall effect (FQHE) in higher LLs. The RIQHE can then be understood as a first melting of an electron solid, upon an increase in the electronic  density (or filling factor), towards a Laughlin liquid followed by an electronic resolidification upon further increase. Qualitatively, this non-monotonic behavior can be understood
from the excitation of quasiparticles (or -holes) once the partial filling factor does not match the ``magical'' values $\bar{\nu}=1/(2s+1)$. Here, $\bar{\nu}$ denotes the filling
factor of the last (partially filled) Landau-level subbranch. These quasiparticles yield an additional contribution to the energy of the Laughlin liquid, such that the total energy 
of the liquid states increases faster than the variation of the electron-solid energies and thus causes the electronic resolidification. 

In the following, we concentrate on the graphene LLs $n=1$ (filling factor $2\leq \nu\leq 6$) and $n=2$ (filling factor $6\leq \nu\leq 10$. 

From a theoretical point of view, electrons in a single LL are described by the Hamiltonian

\begin{equation}\label{eq:ham}
 H=\frac{1}{2} \sum_{\bq} v(q) \left[F_n(\bq)\right]^2\rhobar(-\bq)\rhobar(\bq),
\end{equation}

\noindent where $\rhobar(\bq)$ denotes the projected density operator, while the LL form factor $F_n(\bq)$, which stems from the wavefunction overlap in the $n$-th LL, can be absorbed 
in an \textit{effective} interaction potential $v_n(q) = v(q) [F_n(\bq)]^2$. One notices that the LL form factor also encodes the underlying relativistic or non-relativistic 
nature of the electron. For relativistic electrons in graphene, it reads (for $n\geq 1$) \cite{Goerbig2011grapheneRIQHE}

\begin{equation}
 F_n(\bq) = \frac{1}{2}\left[ L_{n-1}\left(\frac{q^2l_B^2}{2}\right) + L_{n}\left(\frac{q^2l_B^2}{2}\right)\right] e^{-q^2l_B^2/4},
\end{equation}

\noindent in terms of the magnetic length $l_B=\sqrt{\hbar/eB}$ and the Laguerre polynomials $L_n(x)$. 

\begin{figure}
  \centering
  \includegraphics[width=0.5\textwidth]{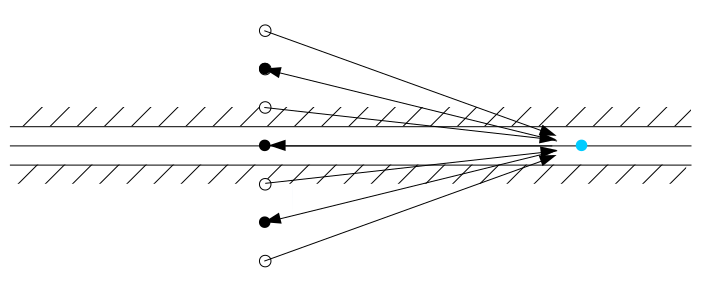}\\
  \caption{ {\bf The screening effect by image charges.} The graphene sheet is represented by the central line, separated by a distance $d$ from the metallic top and bottom gates. 
  The interaction potential felt by the test charge (gray dot) does not only contain the original charge ($e$, black dot) in the graphene sheet, but also its recursive images 
  [succession of white (charge $-e$) and black dots (charge $e$)].}\label{scpoten}
\end{figure}

While in many theoretical approaches, the interaction $v(q)$ is taken to be the bare Coulomb interaction $v(q)= 2\pi e^2/\epsilon q$ (in reciprocal space), in the experimental
sitation here we need to take into account strong screening effects due to the metallic back and top gates. The situation is depicted in Fig. \ref{scpoten}, where a test charge
in the graphene sheet is not only submitted to an original charge in the graphene sheet but also to its mirror images generated by the metallic gates, separated from the graphene sheet
by a distance $d$ that we have chosen here to be the same for the back and top gates. The summation of all
contributions yields an additional factor (in reciprocal space) 

\begin{equation}\label{eq:form}
f(q) = \frac{1+e^{-2dq}-2e^{-dq}}{1-e^{-2dq}},
\end{equation}

\noindent  to the effective interaction potential, which then reads

\begin{widetext}
\begin{equation}
 v_n(q) = \frac{2\pi e^2}{\epsilon q} f(q) \left[F_n(q)\right]^2= \frac{\pi e^2}{2\epsilon q}\frac{1+e^{-2dq}-2e^{-dq}}{1-e^{-2dq}}
 \left[ L_{n-1}\left(\frac{q^2l_B^2}{2}\right) + L_{n}\left(\frac{q^2l_B^2}{2}\right)\right]^2 e^{-q^2l_B^2/2}
\end{equation}
\end{widetext}

In order to describe the different electron-solid phases, we use the Hartree-Fock approximation, which amounts to considering the order parameter 
$\Delta(\bq)=\langle \rhobar (\bq)\rangle/n_B A$, in terms of the flux density $n_B=1/2\pi l_B^2=eB/h$ and the total area $A$ \cite{goerbig2004competition}. The order parameter
can then be understood as the reciprocal-space form of the (guiding-center) density modulation $\nubar(\br)$. In the present note, we explore two complementary 
modulations for the bubble crystals. The first one treats a bubble with $M$ electrons as a symmetric cylinder (of height) one occupying a surface $2\pi M l_B^2$, while the 
cylinders are arranged in a triangular lattice with lattic vectors $\bR_j$

\begin{equation}\label{eq:B1}
\begin{split}
\nubar_M^{0}(\br) = \sum_{\bR_j} \theta(\sqrt{2M}l_B - |\br -\bR_j|), \\
 \Delta_M^0(\bq)= \frac{2\pi \sqrt{2M}l_B}{A q}J_1(\sqrt{2M}ql_B)\sum_{\bR_j}e^{i\bq\cdot\bR_j},
\end{split}
\end{equation}

\noindent  in terms of the first-order Bessel function $J_1(x)$. The second approach is more sophisticated and more accurate -- it takes into account quantum correlations inside each bubble, since the electrons have a tendency to avoid each other due to the Pauli principle. In this case, each bubble is considered as a Laughlin droplet 
$\prod_{1\leq j\leq k\leq M}(z_j - z_k) \exp( -\sum_{j=1}^M|z_j|^2/4l_B^2)$ and the order parameter reads \cite{Folger1997}

\begin{equation}\label{eq:B2}
 \Delta_M(\bq)= \frac{2\pi l_B^2}{M A}L_{M-1}^1\left(\frac{q^2l_B^2}{2}\right)e^{-q^2l_B^2/4}\sum_{\bR_j}e^{i\bq\cdot\bR_j}.
\end{equation}

Notice that both formulations are asymptotically identical in the large-size limit ($M\gg 1$).

With the help of these order parameters, the Hartree-Fock energy reads 

\begin{equation}\label{eq:HF}
 E_B=\frac{n_B}{2\nubar} \sum_{\bq} u_n^{HF}(q)|\Delta_M (q)|^2, 
\end{equation}

\noindent in terms of the Hartree-Fock potential $u_n^{HF}(q)= v_n(q) - \sum_{\bp} v_n(p) \exp[i(q_xp_y - q_yp_x)l_B^2]/n_B A$. 

This energy needs to be compared to those of the Laughlin states, and their energy at $\nubar=1/(2s+1)$ can be written as 

\begin{equation}\label{eq:LL}
 E_L(s) = \frac{\nubar}{\pi} \sum_{m=0} c_{2m+1}^s V_{2m+1}^n, 
\end{equation}

\noindent in terms of Haldane's pseudopotentials \cite{Haldane1983} $V_{2s+1}^n= (2\pi/A)\sum_{\bq} v_n(q) L_{2m+1}(q^2l_B^2)\exp(-q^2l_B^2/2)$. The coefficients characterize entirely 
the Laughlin liquid and are subject to sum rules \cite{Levesque1984,Girvin1986} that can be used as a system of linear equations to obtain these coefficients \cite{goerbig2004competition}. This procedure 
reproduces to great accuracy the energy of the Laughlin liquids. 

Equations (\ref{eq:HF}) and (\ref{eq:LL}) allow us to compare the energies of the different phases as a function of the partial filling factor $\nubar$ and the gate distance $d$. 
Since there is no evidence for stripe phases, their energy is not calculated although their description is not difficult. 

\begin{figure}[H]
\centering
\includegraphics[width=0.45\textwidth]{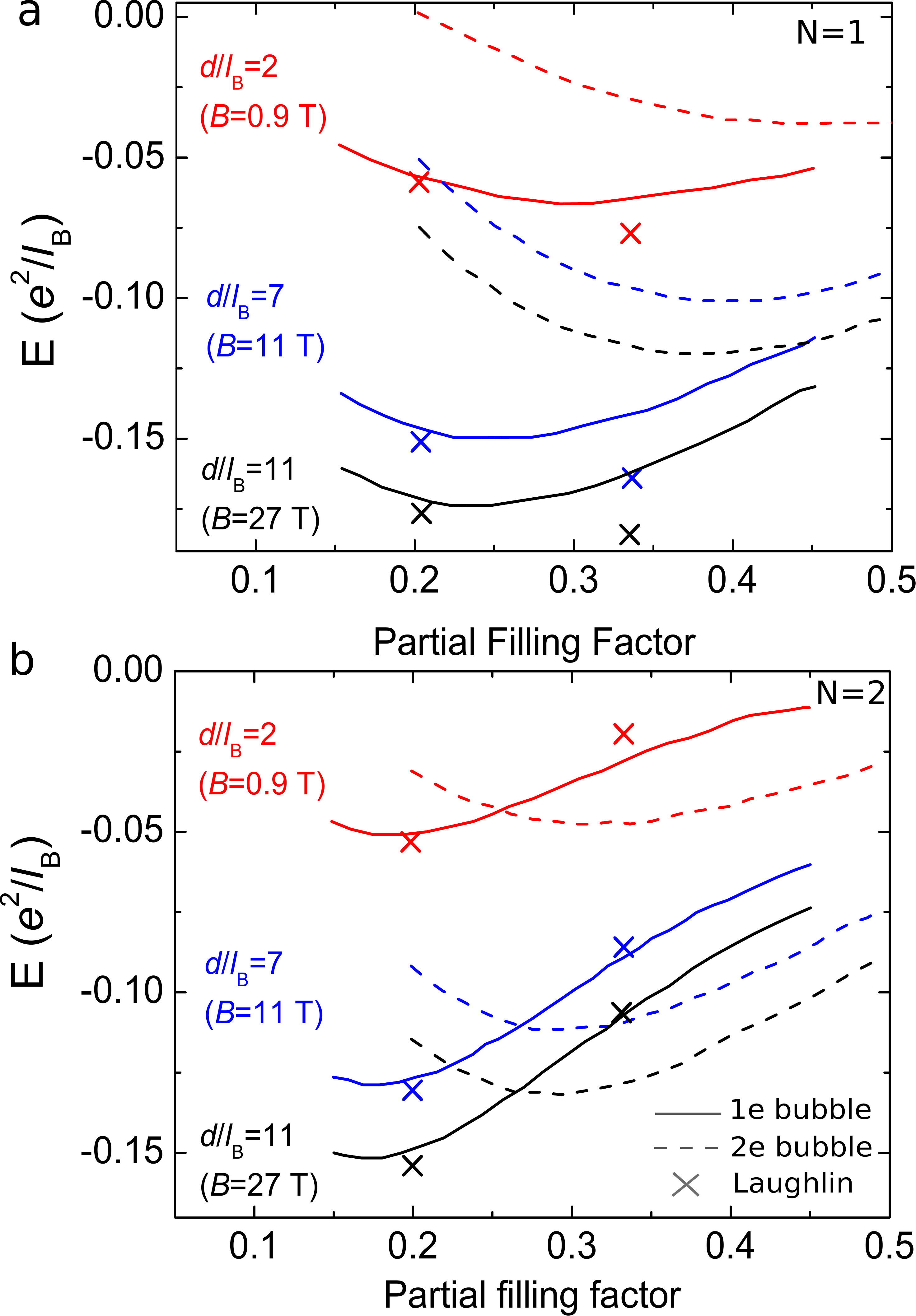} 
\caption{{\bf Energy comparison curves for different  $d / l_{\rm B}$ parameter}. Assuming an approximate symmetry $\nubar \rightarrow 1-\nubar$ within each LL subbranch,
we only show the energy curves in a range $0\leq \nubar\leq 1/2$. Left panel N$=$1, rigth panel for N$=$2. The energy is plotted in units of $e^2/\epsilon l_B$. The crosses indicate the energy of the Laughlin liquids, while the continuous lines show that of the 
bubble crystals with $M=1$ (Wigner crystal) and $M=2$ electrons per bubble. The curves to the right are those of $M=2$. These curves have been obtained for an order
parameter \ref{eq:B2}.}\label{fig:ener}
\end{figure}

The energy curves are plotted in Fig. \ref{fig:ener} for different values of  $d / l_{\rm B}$, i.e. in units of the magnetic length $l_B=26$ nm$/\sqrt{B{\rm [T]}}$. For an experimentally
relevant distance of $d\sim 27$ nm, one has $d / l_{\rm B}\sim 2$ at $B=0.9$ T and $d / l_{\rm B}\sim 7$ at $B=11$ T, corresponding roughly to the red and blue lines in
Fig. \ref{fig:ener}. One notices that in N$=$1, the Laughlin state at $\nubar =1/3$ is always well developed, while that at $\nubar=1/5$ is energetically extremely close, albeit lower,
than the Wigner crystal ($M=1$). The blue lines indicate the result for bubbles with internal Laughlin-type correlations, given by the order parameter (\ref{eq:B2}), while the
green ones show the bubble-phase energies for the order parameter (\ref{eq:B1}). Generically (apart from gates very close to the graphene sheet in N$=$2), bubble crystals with
internal Laughlin-type correlations are favored. However, the overall succession of phases is unaltered by the alternative choice of the order parameter. Based on these results,
we expect in N$=$1 a RIQHE between $\nubar=1/5$ and $\nubar=1/3$. Depending on the quasiparticle energy (not calculated here), there could also exist a RIQHE between $\nubar=1/3$ 
and the nonquantized liquid at $\nubar=1/2$.

The situation is different in $n=2$, where the Laughlin liquid at $\nubar=1/3$ ceases to be the state of lowest energy. Indeed, one obtains a bubble crystal with $M=2$ electrons
per bubble. Notice that the Laughlin liquid has an even higher energy than the Wigner crystal with $M=1$ for all values of $d$. In contrast, the Laughlin liquid at $\nubar=1/5$ 
remains the ground state for all values of $d$, regardless of the ansatz we use for the bubble-crystal order parameter. One therefore expects, upon increase of $\nubar$, first a 
melting of the Wigner crystal to the $\nubar=1/5$ FQHE and then a resolidification causing the RIQHE. Interestingly, for rather large values of $d$ (middle and right columns), 
one would expect first a resolidification to a Wigner crystal that then transits to a $M=2$ bubble crystal, within a first-order phase transition. This is reminiscent to 
GaAs heterostructures \cite{goerbig2004competition}, where the phase coexistence between a Wigner and a bubble crystal has been experimentally proven within microwave experiments 
\cite{Lewis2004}.

\section*{Energy scale in the dual gated calculations}

Following the calculation results, we point out that screening due to the top and back gates does not alter the succession of quantum phases, and our theoretical results thus agree with those by Knoester \textit{et al.}
\cite{knoester2016electron}. However, screening has a drastic influence on the overall energy scale. While the natural energy scale in monolayer graphene is given by 
$e^2/\epsilon l_B$, the presence of the form factor changes the nature of the interaction potential. As one can already appreciate from a single gate, the mirror charge provides
a dipole, and the interaction potential becomes dipolar at long distances. In the case of a single gate, the form factor in Eq. (\ref{eq:form}) would read

\begin{equation}\label{eq:form2}
 f_{dp}(q)=1- e^{-2qd}\sim 2qd, 
\end{equation}

where the last approximation stems from an expansion in the long-distance (small-$q$) limit. The Coulomb interaction thus needs to be replaces by 
$e^2/\epsilon q \rightarrow 2e^2 d/\epsilon$. These rough arguments in orders of magnitude indicate that the overall energy scale is replaced by 

\begin{equation}
 \frac{e^2}{\epsilon l_B}\rightarrow \frac{e^2}{\epsilon l_B}\frac{d}{l_B} \propto d \times B.
\end{equation}

The linear scaling in the distance $d$ of the overall energy scale is quite apparent in our theoretical results in Fig. \ref{fig:ener}, where a variation between $d/l_B=0.5$ and
$5$ yields an energy increase by one order of magnitude. Naturally, one expects the effect to saturate in the large $d/l_B$-limit, where the expansion in Eq. (\ref{eq:form2})
is no longer justified since it corresponds to a large $dq$-limit where the form factor approaches 1, as expected for an unscreened potential. However, in the present intermediate
regime, where this expansion seems rather well justified, one would also expect a scaling of the energy scale that is linear in the magnetic field $B$. This could be used as 
a clear experimental fingerprint of screened interactions, e.g. in measuring activation gaps, since, in the unscreened limit, one expects a scaling in $\sqrt{B}$.

\section*{Magnetic field evolution of R6a}
\begin{figure}[H]
\centering
\includegraphics[width=1\linewidth]{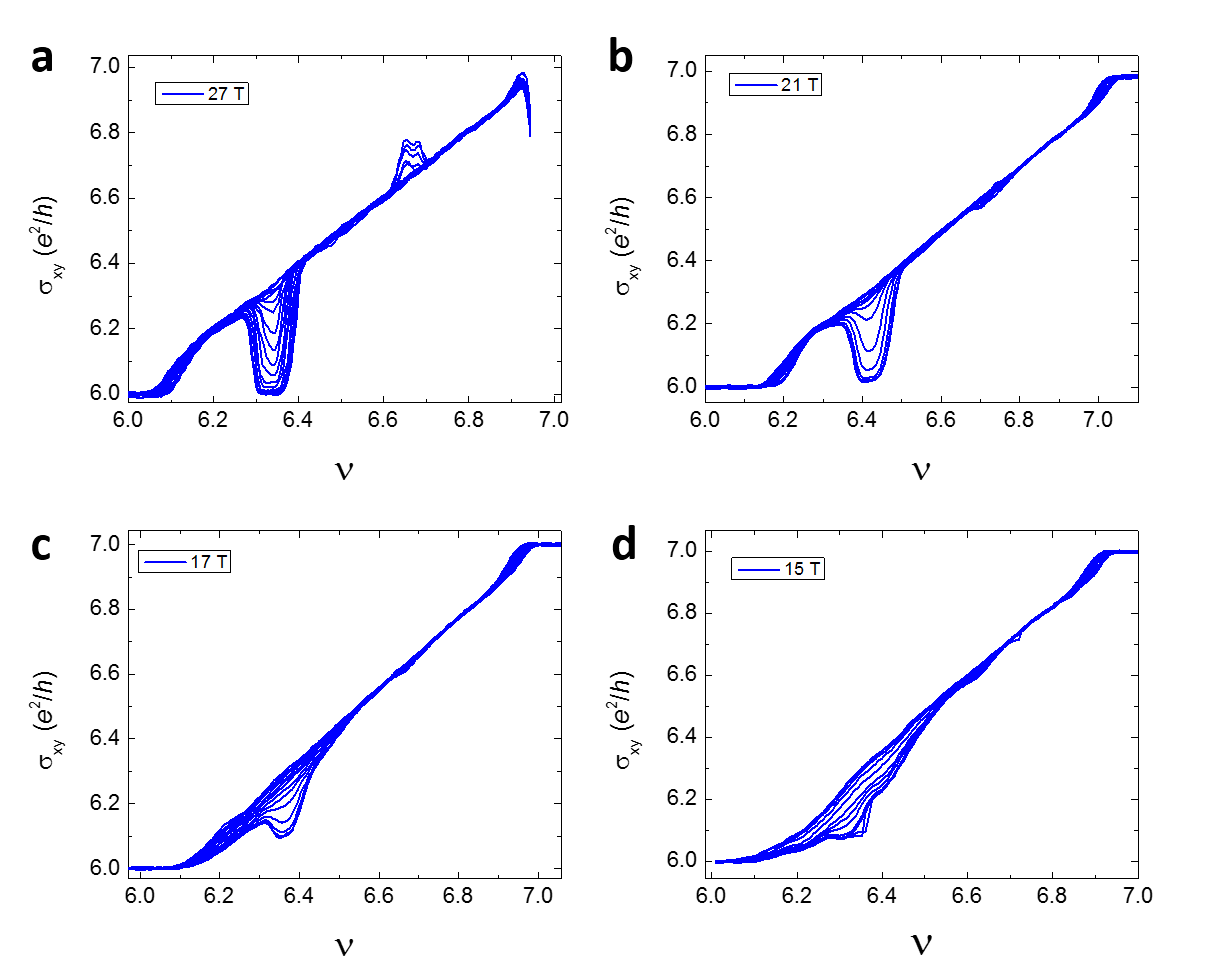} 
\caption{{\bf Magnetic field evlution}. Temperature dependence for different magnetic field: 27 T{\bf a}, 21 T {\bf b}, 17 T {\bf c} and 15 T {\bf b}. Temperature range from 0.3 K to 1.2 K. }
\label{fig:figS_temp}
\end{figure}

\section*{Thermal cycle curves}

\begin{figure}[H]
\centering
\includegraphics[scale=0.5]{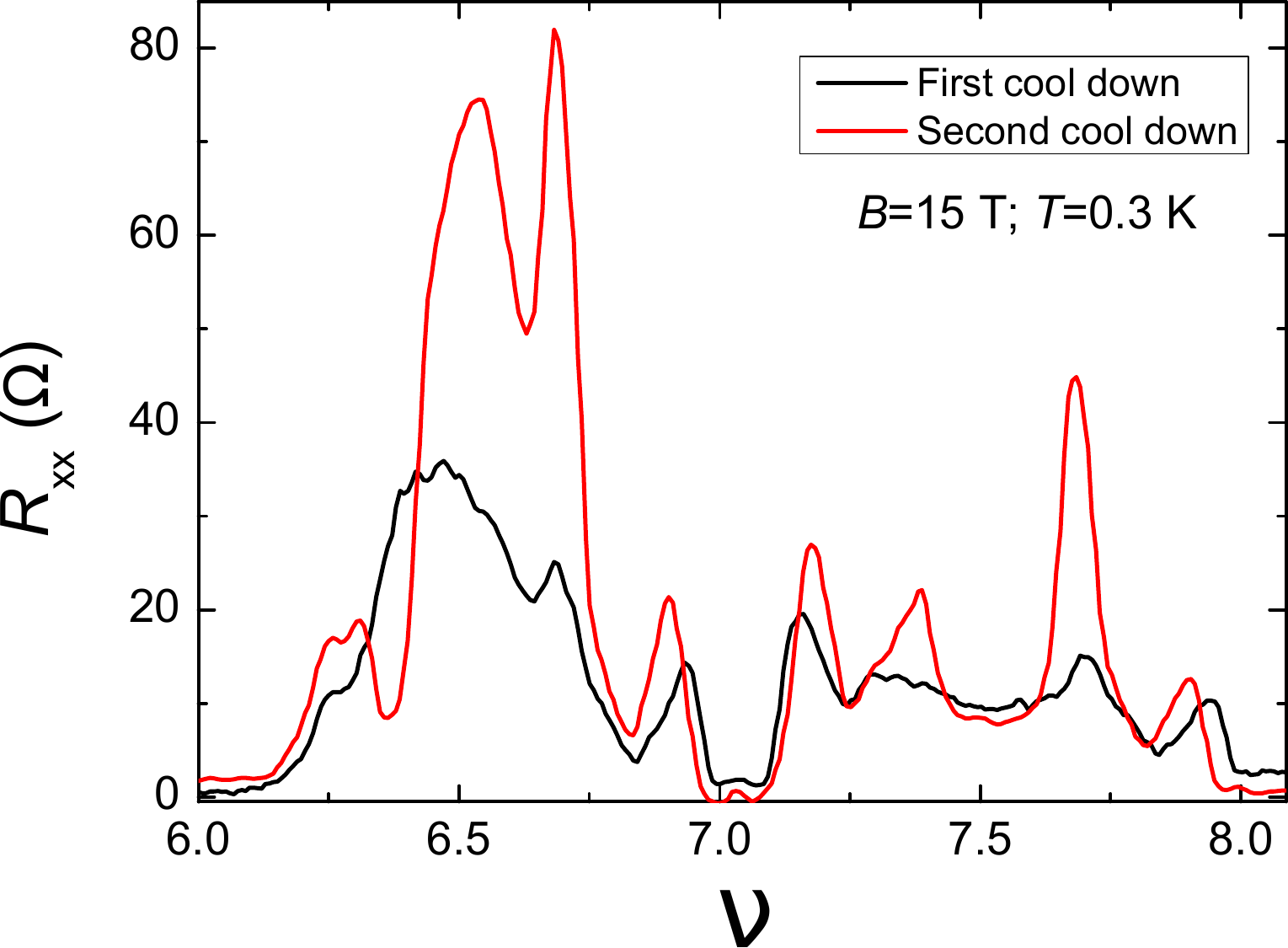} 
\caption{{\bf $R_{\textrm{xx}}$ for two different cool downs.} Longitudinal resistance of the same sample and the same measurement configuration for two different cool downs. Notice that the base temperature 0.3 K and magnetic field 15 T of both measurements are the same.}
\label{fig:figS_thermal}
\end{figure}

\section*{Transverse measurements}

\begin{figure}[H]
\centering
\includegraphics[scale=0.3]{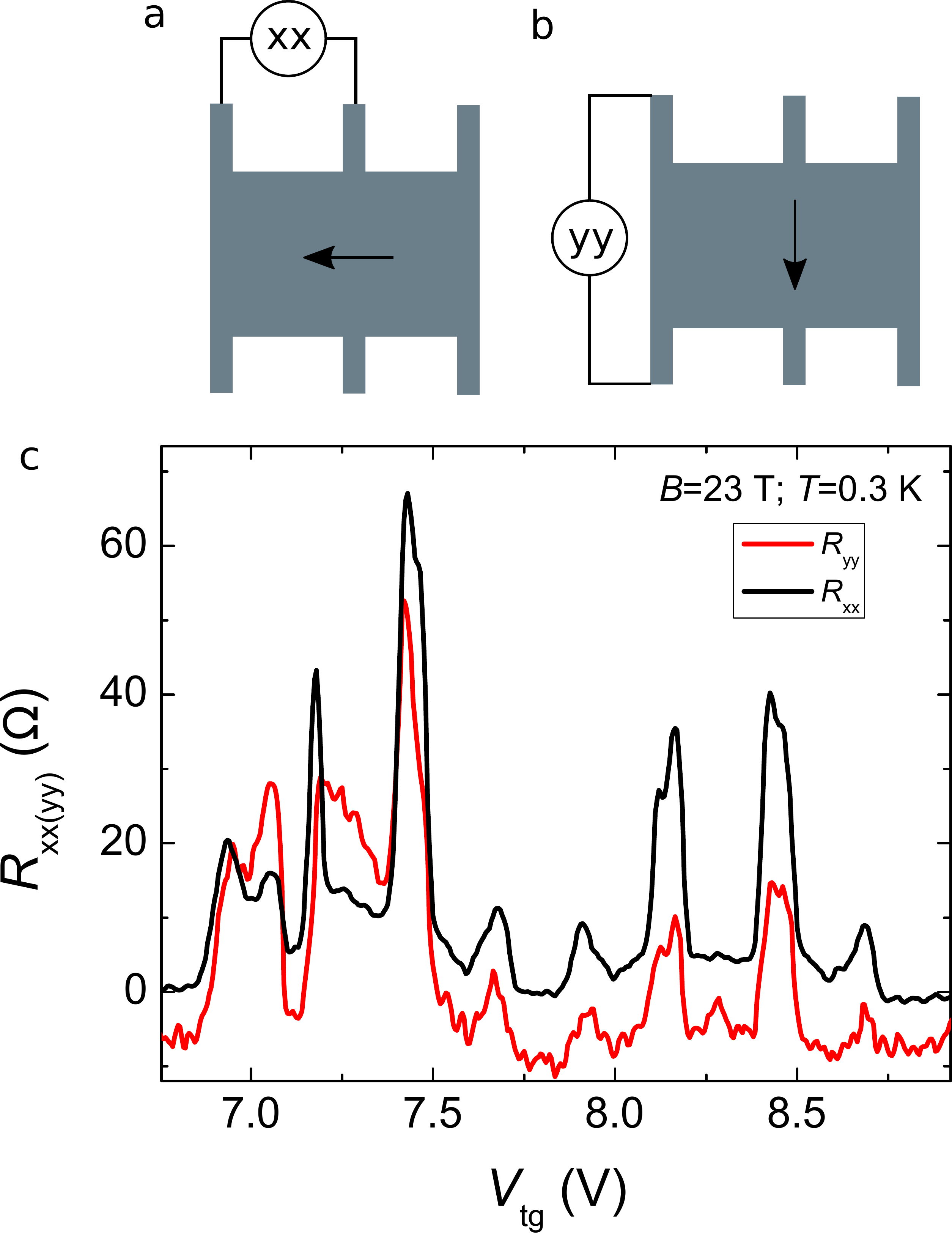} 
\caption{{\bf Longitudinal conductance along two directions.} {\bf a} and {\bf b}, sketch of the two measurement configurations, separated 90 degrees from each other. {\bf c} longitudinal resistance for the measurement configurations depicted in {\bf a} and {\bf b}. Measured at 23 T and 0.3 K.}
\label{fig:figS_trans}
\end{figure}


\end{document}